\documentclass[12pt]{article}

\usepackage[fleqn]{amsmath}
\usepackage{amssymb} 
\usepackage{hhline}
\usepackage[scale={.75,.75}]{geometry}
\usepackage{cite}
\usepackage{epsfig}

\newcommand{\br}{\mathrm{BR}}

\newcommand{\mplanck}{M_{\text{Pl}}}

\newcommand{\gsim}{\lower.7ex\hbox{$\;\stackrel{\textstyle>}{\sim}\;$}}
\newcommand{\lsim}{\lower.7ex\hbox{$\;\stackrel{\textstyle<}{\sim}\;$}}

\numberwithin{equation}{section}
\numberwithin{table}{section}
\allowdisplaybreaks

\begin{document}
\date{\mbox{ }}

\title{ 
{\normalsize     
DESY 09-055\hfill\mbox{}\\
TUM-HEP 720/09\hfill\mbox{}\\
June 2009\hfill\mbox{}\\}
\vspace{1cm}
\bf Probing Gravitino Dark Matter\\
with PAMELA and Fermi\\[8mm]}
%
\author{Wilfried~Buchm\"uller$^a$, Alejandro Ibarra$^b$,
Tetsuo Shindou$^c$,\\ Fumihiro Takayama$^a$ and David Tran$^b$\\[2mm]
{\normalsize\it $^a$Deutsches Elektronen-Synchrotron DESY, 22607 Hamburg, 
Germany}\\[2mm]
{\normalsize\it $^b$Physik-Department T30d, Technische Universit\"at M\"unchen,
85748 Garching, Germany}\\[2mm]
{\normalsize\it $^c$Kogakuin University, 1638677 Tokyo, Japan}
}
\maketitle

\thispagestyle{empty}

\begin{abstract}
\noindent
We analyse the cosmic-ray signatures of decaying gravitino dark matter in a 
model-independent way based on an operator analysis. Thermal leptogenesis 
and universal boundary conditions at the GUT scale restrict the gravitino
mass to be below 600~GeV. Electron and positron fluxes from gravitino
decays, together with the standard GALPROP background, cannot explain both,
the PAMELA positron fraction and the electron $+$ positron flux recently
measured by Fermi LAT. For gravitino dark matter, the observed fluxes
require astrophysical sources. The measured antiproton flux allows for
a sizable contribution of decaying gravitinos to the gamma-ray 
spectrum, in particular a line at an energy below 300~GeV. Future measurements 
of the gamma-ray flux will provide important constraints on possible signatures
of decaying gravitino dark matter at the LHC. 
\end{abstract}

\newpage

\maketitle

\section{Introduction}

An unequivocal prediction of locally supersymmetric extensions of the Standard 
Model is the gravitino, the gauge fermion of supergravity 
\cite{fnf76}. Depending on the mechanism of supersymmetry breaking, it can be 
the lightest superparticle, which makes it a natural dark matter candidate 
\cite{pp81}. In connection with thermal leptogenesis \cite{fy86}, gravitino
dark matter has been discussed as an alternative \cite{bbp98} to the standard 
WIMP scenario \cite{fen05}. 

In a class of models with small $R$-parity and lepton number breaking the 
gravitino is no longer stable, but its decays into Standard Model (SM) 
particles are doubly suppressed by the Planck mass and the small $R$-parity 
breaking parameter. Hence, its lifetime can exceed the age of the Universe 
by many orders of magnitude, and the gravitino remains a viable dark matter 
candidate \cite{ty00}. Recently, it has been shown that such models yield a
consistent cosmology incorporating nucleosynthesis, leptogenesis and gravitino
dark matter \cite{bcx07}.

Small $R$-parity breaking can arise from spontaneous B-L breaking \cite{bcx07} 
or from left-right symmetry breaking \cite{jmn08}. Alternatively, explicit
$R$-parity violating couplings of heavy Majorana neutrinos can lead to suppressed $R$-parity 
breaking interactions in the low energy effective theory via the seesaw
mechanism \cite{es09}. In the simplest supergravity models with universal 
gaugino masses at the Grand Unification (GUT) scale, thermal leptogenesis 
implies an
upper bound of 600~GeV on the gravitino mass \cite{bes08}. Relaxing the
boundary conditions at the GUT scale, gravitino masses up to 1.4~TeV are
possible \cite{hty09}.

Gravitino decays may lead to characteristic signatures in high-energy cosmic 
rays. 
The produced flux of gamma rays \cite{ty00,bcx07,lor07,bbx07,it07,imm08}
and positrons \cite{it08,imm08} has been found to potentially account 
for the extragalactic component
of the excess in the EGRET \cite{smr04} and HEAT \cite{heat97} data, 
respectively. Furthermore, a neutrino flux from gravitino decays is
predicted \cite{cgx08} as well as a possibly observable antideuteron
flux \cite{Ibarra:2009tn}.

Recently, a steep rise in the cosmic-ray positron fraction above 10~GeV has
been discovered by the PAMELA collaboration \cite{pamela08} whereas the 
observed antiproton-to-proton ratio \cite{pamela'08} is consistent with 
previous measurements of the antiproton flux by BESS~\cite{Orito:1999re}, 
IMAX~\cite{Mitchell:1996bi} and WiZard/CAPRICE~\cite{Boezio:1997ec}. 
A possible explanation of this exotic positron source is annihilating 
or decaying dark matter \cite{ber09}, including decaying 
gravitinos \cite{imm'08,it'08}. Equally important are the recent
measurements of the total electron $+$ positron flux by ATIC \cite{atic08},
H.E.S.S. \cite{hess08} and Fermi LAT \cite{fermi09}.  

In this paper we analyse the cosmic-ray signatures of decaying gravitino
dark matter in a model-independent way based on an operator analysis.
Consistency with the observed antiproton flux yields a
lower bound on the gravitino lifetime. As we shall see, this determines
an upper bound on the continuous gamma-ray spectrum.
Following a previous analysis of supergravity models and leptogensis, we only
consider gravitino masses below 600~GeV. Hence, gravitino decays
cannot be the cause of the anomaly observed by the ATIC \cite{atic08}
and Fermi LAT \cite{fermi09} collaborations.
An interpretation of the PAMELA positron anomaly as 
the result of gravitino dark matter decay, 
on the other hand, requires gravitino masses above 200~GeV.

This paper is organised as follows. In Section~2 we present a general
operator analysis of gravitino decays and study the implications for
the different branching ratios. In particular we discuss the strength of
the predicted monochromatic line in the gamma-ray spectrum. Section~3 
deals with the electron, positron and antiproton flux from gravitino decays
and the implications for the gravitino lifetime.
The results for the gamma-ray spectrum are discussed in Section~4, 
followed by our conclusions in Section~5.\\

\section{Gravitino decays}

$R$-parity violating gravitino decays are conveniently described in terms of
effective operators. The spinor of a massive gravitino (cf.~\cite{wb92}) 
satisfies the Dirac equation
\begin{align}\label{dirac}
\left(i\gamma^\mu \partial_\mu - m_{3/2}\right)\psi_{\nu} = 0\ ,
\end{align}
together with the constraints
\begin{align}\label{constraints}
\gamma^{\mu} \psi_{\mu} = 0\ ,\quad
\partial^{\mu}\psi_{\mu} = 0\ .
\end{align}
The mass scales multiplying the non-renormalizable operators are inverse powers
of the Planck mass $\mplanck$ and the supersymmetry breaking gravitino mass
$m_{3/2}$. This assumes for the masses $m_{\text{SM}}$ of Standard Model 
particles, the gravitino mass and the masses $m_{\text{soft}}$ of other
superparticles the hierarchy 
$m^2_{\text{SM}} \ll m^2_{3/2} \ll m^2_{\text{soft}}$.

Using Eqs.~(\ref{dirac}) and (\ref{constraints}) one easily verifies that
the dimension-5 and dimension-6 operators for $R$-parity violating
couplings of the gravitino to Standard Model particles are given by
\begin{align}\label{eff}
{\cal L}_{\mathrm{eff}} = 
\frac{i\kappa}{\sqrt{2}\mplanck}
\left\{\bar{l}\gamma^\lambda\gamma^\nu D_\nu \phi \psi_{\lambda} 
+\frac{i}{2}\bar{l}\gamma^\lambda
\left(\xi_1 g'YB_{\mu\nu} + \xi_2 g W_{\mu\nu}\right)
\sigma^{\mu\nu}\phi\psi_{\lambda} \right\} + \mathrm{h.c.} \ ,
\end{align}
where typically $\xi_{1.2} = {\cal O}(1/m_{3/2})$. Note, however, that
in general $\kappa$ and the product $\kappa \xi_{1,2}$ are independent 
parameters. For simplicity, we have suppressed
the flavour indices of $\kappa$, $\l$ and $\xi_{1,2}$.
The covariant derivative involves
the $\mathrm{U(1)}$ and $\mathrm{SU(2)}$ gauge fields $B_\mu$ and 
$W_\mu$, respectively, 
\begin{align}
D_\mu = \partial_\mu + ig' Y B_\mu + i g W_\mu \ ,\quad 
Y[\phi] = -\frac{1}{2}\ ,\quad
W_\mu = \frac{1}{2} \sigma^I W_\mu^I\ ,
\end{align}
with the corresponding field strengths 
\begin{align}
B_{\mu\nu} = \partial_\mu B_\nu - \partial_\nu B_\mu\ ,\quad
W_{\mu\nu} = \partial_\mu W_\nu - \partial_\nu W_\mu\ + i[W_\mu,W_\nu]\ .
\end{align}
In the unitary gauge, the Higgs and lepton doublets read
\begin{align}
\phi = \begin{pmatrix} v + \frac{1}{\sqrt{2}} h \\ 0 \end{pmatrix}\ ,\quad
l = \begin{pmatrix} \nu \\ e \end{pmatrix}\ ,
\end{align}
where $h$, $\nu$ and $e$ denote Higgs boson, neutrino and charged lepton,
respectively.

From Eq.~(\ref{eff}) one easily obtains the couplings of leptons and gravitino
to Higgs and gauge bosons, which are responsible for the two-body gravitino
decays,
\begin{align}\label{cubic}
{\cal L}_3 \supset 
\frac{i\kappa}{\sqrt{2}\mplanck}
\{&\left(\partial_\mu h + im_Z Z_\mu\right)
\bar{\nu}\gamma^\nu\gamma^\mu \psi_{\nu} 
+ i \sqrt{2} m_W W^-_\mu \bar{e}\gamma^\nu\gamma^\mu \psi_{\nu}
\nonumber\\ 
& + i m_Z\left(\xi_Z \partial_\mu Z_\nu + \xi_\gamma \partial_\mu A_\nu\right)
\bar{\nu}\gamma^\lambda \sigma^{\mu\nu} \psi_{\lambda} \nonumber\\
& + i \sqrt{2} m_W\xi_W \partial_\mu W^-_\nu 
\bar{e}\gamma^\lambda\sigma^{\mu\nu} \psi_{\lambda} \} + \mathrm{h.c.} \ ,
\end{align}
with
\begin{align}
\xi_Z &= \sin^2{\theta_{\rm W}}\xi_1 + \cos^2{\theta_{\rm W}}\xi_2\ ,\quad
\quad \xi_W = \xi_2\ ,\nonumber\\
\xi_\gamma &= \sin{\theta_{\rm W}}\cos{\theta_{\rm W}}
\left(\xi_2 - \xi_1\right)\ ,
\quad \sin{\theta_{\rm W}} = \frac{g'}{\sqrt{g^{'2}+g^2}}\ . \label{xi}
\end{align}
Note that the gauge boson couplings satisfy the relation
\begin{align}
\xi_Z + \tan{\theta_{\rm W}} \xi_\gamma = \xi_W\ .
\end{align}
For $\xi_1 = \xi_2$, one has $\xi_\gamma = 0$ and $\xi_Z = \xi_W$.

The interaction Lagrangian (\ref{cubic}) coincides with the one obtained
from bilinear $R$-parity breaking \cite{cgx08,gre08} if parameters are properly 
matched\footnote{
The relations are $\kappa_i = \langle \widetilde{\nu}_i \rangle/v$, 
$U_{\widetilde{Z}\widetilde{Z}}=-m_Z\xi_Z$, 
$U_{\widetilde{W}\widetilde{W}}=-m_W\xi_W$, 
$U_{\widetilde{\gamma}\widetilde{Z}}=-m_Z\xi_\gamma$,
$U_{\widetilde{H}_u\widetilde{Z}}\sin{\beta} +
U_{\widetilde{H}_d\widetilde{Z}}\cos{\beta} 
+ m^2_{\widetilde{\nu}_\tau}/(m^2_{\widetilde{\nu}_\tau} - m^2_h) = 1$.}. 
Using the results of \cite{cgx08,gre08} we then obtain the partial 
gravitino decay widths
\begin{align}
\Gamma(\psi_{3/2}\rightarrow h\nu_i) &=
\frac{\kappa_i m_{3/2}^3}{384\pi\mplanck^2}\beta_h^4\ ,\label{bh}\\
\Gamma(\psi_{3/2}\rightarrow \gamma\nu_i) &=
\frac{\kappa_i |\xi_{\gamma i}|^2 m_Z^2 m_{3/2}^3}{64\pi\mplanck^2}\ ,
\label{bgamma}\\
\Gamma(\psi_{3/2}\rightarrow Z\nu_i) &=
\frac{\kappa_i m_{3/2}^3}{384\pi\mplanck^2}\beta_Z^2
\left(H_Z + 16 \frac{m_Z^2 {\rm Re}(\xi_{Z i})}{m_{3/2}}G_Z 
+ 6 m_Z^2|\xi_{Z i}|^2 F_Z\right)\ , \label{bz}\\
\Gamma(\psi_{3/2}\rightarrow W^{\pm}e^{\mp}_i) &=
\frac{\kappa_i m_{3/2}^3}{192\pi\mplanck^2}\beta_Z^2
\left(H_W + 16 \frac{m_W^2 {\rm Re}(\xi_{W i})}{m_{3/2}}G_W 
+ 6 m_W^2|\xi_{W i}|^2 F_W\right)\label{bw}\ ,
\end{align}
where the subscript $i$ denotes the generation index. The functions
$\beta_a$, $H_a$, $G_a$ and $F_a$ ($a=h,Z, W$) 
are given by \cite{cgx08}
\begin{align}
\beta_a &= 1 - \frac{M_a^2}{m_{3/2}^2}\ ,\\
H_a &= 1 + 10 \frac{M_a^2}{m_{3/2}^2}
+ \frac{M_a^4}{m_{3/2}^4}\ ,\\
G_a &= 1 + \frac{1}{2} \frac{M_a^2}{m_{3/2}^2}\ ,\\
F_a &= 1 + \frac{2}{3} \frac{M_a^2}{m_{3/2}^2}
+ \frac{1}{3}\frac{M_a^4}{m_{3/2}^4}\ .
\end{align}
As expected, one has for $m_a\xi_a \sim m_a/m_{3/2} \ll 1$,
\begin{align}
\Gamma(\psi_{3/2}\rightarrow h\nu_i) &\simeq
\Gamma(\psi_{3/2}\rightarrow Z\nu_i) \simeq
\frac{1}{2}\Gamma(\psi_{3/2}\rightarrow W^{\pm}e^{\mp}_i)\ .
\end{align}

The decay width $\Gamma(\psi_{3/2}\rightarrow \gamma\nu)$ is of particular
interest since it determines the strength of the gamma line at the end of
the continuous spectrum. As discussed above, this decay width is 
model-dependent. Contrary to the continuous part of the spectrum it can vanish,
which is the case for $\xi_1 - \xi_2 = 0$. Generically, without such a 
cancellation, one obtains for the branching ratio using Eqs.~(\ref{xi})
and (\ref{bh}) -- (\ref{bw}):
\begin{equation}\label{line}
\br(\psi_{3/2}\rightarrow \gamma\nu) 
\sim 0.3 \sin^2{\theta_W} \left(\frac{m_Z}{m_{3/2}}\right)^2
\sim 0.02 \left(\frac{200~\mathrm{GeV}}{m_{3/2}}\right)^2 \ .
\end{equation}
This estimate will be used in Section~4 where the gamma-ray spectrum is
discussed.

Another phenomenologically important issue is the flavour structure of
gravitino decays, i.e., the dependence of the parameters $\xi_{\gamma i}$,
$\xi_{Z i}$ and $\xi_{W i}$ on the generation index. In models of bilinear
$R$-parity breaking, for instance, this information is encoded in the 
mixing parameters $\mu_i$,
\begin{equation}
\Delta \mathcal{L} = \mu_i H_u L_i\ ,
\end{equation}
where $H_u$ and $L_i$ are Higgs and lepton-doublet superfields, respectively.
In models of flavour the ratios of the parameters $\mu_i$ are related to
the structure of the Majorana neutrino mass matrix,
\begin{equation}
M_\nu = c_{ij} (l_i\phi)(l_j\phi)\ .
\end{equation}
An interesting example, which can account for the large mixing angles in
the neutrino sector, is `anarchy' \cite{hmw99} where $c_{ij} = \mathcal{O}(1)$.
In this case one also has
\begin{equation}\label{anarchy} 
\frac{\mu_i}{\mu_j} = \mathcal{O}(1)\ .
\end{equation}
For the purpose of illustration we shall use in the following sections the
`democratic' case $\mu_1 = \mu_2 = \mu_3$ where the parameters 
$\xi_{\gamma i}$, $\xi_{Z i}$ and $\xi_{W i}$ have no flavour dependence.
Alternatively, one may consider `semi-anarchy' (cf.~\cite{bcx07}). Other 
examples can be found in \cite{af04}.

In the following we shall consider gravitino masses between 100~GeV and
600~GeV, for which the assumed hierarchy 
$m^2_{\text{SM}} \ll m^2_{3/2} \ll m^2_{\text{soft}}$
can only be a rough approximation. A more detailed treatment would have
to incorporate mixings with heavy particles of the supersymmetric
standard model. However, we find the operator analysis useful to illustrate
the main qualitative features of gravitino decays.

\section{Antimatter from gravitino decays}

The scenario of decaying gravitino dark matter provides, 
for a wide range of gravitino masses and lifetimes,
a consistent thermal history of the Universe, incorporating
successful primordial nucleosynthesis and successful baryogenesis 
through leptogenesis. Furthermore, if dark matter
gravitinos decay at a sufficiently large rate, the decay products
could be detected through an anomalous contribution to the
high-energy cosmic ray fluxes. In this section we shall discuss
the constraints on the gravitino parameters which follow from
the observations of the positron fraction by HEAT and PAMELA and 
of the antiproton flux by BESS, IMAX and WiZard/CAPRICE. 

The rate of antimatter production per unit energy and
unit volume at the position $\vec{r}$ with respect to the center of 
the Milky Way is given by
\begin{equation}
Q(E,\vec{r})=\frac{\rho(\vec{r})}{m_{3/2}\tau_{3/2}}\frac{dN}{dE}\;,
\label{source-term}
\end{equation}
where $dN/dE$ is the energy spectrum of antiparticles produced in the 
decay, which we calculated employing event generator 
PYTHIA 6.4~\cite{Sjostrand:2006za}. On the other hand, $\rho(\vec{r})$
is the density profile of gravitinos in the Milky Way
halo. For definiteness we shall adopt the spherically symmetric 
Navarro-Frenk-White halo density profile~\cite{Navarro:1995iw}:
\begin{equation}
\rho(r)=\frac{\rho_0}{(r/r_c)
[1+(r/r_c)]^2}\;,
\end{equation}
with $\rho_0\simeq 0.26\,{\rm GeV}/{\rm cm}^3$ and $r_c\simeq 20 ~\rm{kpc}$,
although our conclusions are not very sensitive to the choice of the density profile.

After being produced in the Milky Way halo, charged cosmic rays
propagate in the Galaxy and its vicinity in a rather complicated way
before reaching the Earth.
Antimatter propagation in the Milky Way is commonly described by
a stationary two-zone diffusion model with cylindrical boundary 
conditions~\cite{ACR}. Under this approximation, 
the number density of antiparticles
per unit kinetic energy, $f(T,\vec{r},t)$, satisfies the following
transport equation, which applies for both positrons and antiprotons:
\begin{equation}
0=\frac{\partial f}{\partial t}=
\nabla \cdot [K(T,\vec{r})\nabla f] +
\frac{\partial}{\partial T} [b(T,\vec{r}) f]
-\nabla \cdot [\vec{V_c}(\vec{r})  f]
-2 h \delta(z) \Gamma_{\rm ann} f+Q(T,\vec{r}) ,
\label{transport}
\end{equation}
where reacceleration effects and non-annihilating
interactions of antimatter in the Galactic disk 
have been neglected, since the primary particles produced by the
gravitino decay rarely cross the disk before reaching the Earth.

The first term on the right-hand side of the transport equation
is the diffusion
term, which accounts for the propagation through the
tangled Galactic magnetic field.
The diffusion coefficient $K(T,\vec{r})$ is assumed to be constant
throughout the diffusion zone and is parametrised by:
\begin{equation}
K(T)=K_0 \;\beta\; {\cal R}^\delta ,
\end{equation}
where
$\beta=v/c$ and ${\cal R}$ is the rigidity of the particle, which
is defined as the momentum in GeV per unit charge, 
${\cal R}\equiv p({\rm GeV})/Z$.
The normalization $K_0$ and the spectral index $\delta$
of the diffusion coefficient are related to the properties 
of the interstellar medium and can be determined from the 
flux measurements of other cosmic ray species, mainly from 
the Boron-to-Carbon (B/C) ratio~\cite{Maurin:2001sj}. 
The second term accounts for energy losses due to 
inverse Compton scattering on starlight or the cosmic microwave 
background, synchrotron radiation and ionization. 
The third term is the convection term, which accounts for
the drift of charged particles away from the 
disk induced by the Milky Way's Galactic wind. 
It has axial direction and is also assumed to be constant
inside the diffusion region: 
$\vec{V}_c(\vec{r})=V_c\; {\rm sign}(z)\; \vec{k}$.
The fourth term accounts for antimatter annihilation with rate
$\Gamma_{\rm ann}$, when it interacts with ordinary matter
in the Galactic disk,
which is assumed to be an infinitely thin disk with half-width
$h=100$ pc. 
Lastly, $Q(T,\vec{r})$ is the source term of positrons or
antiprotons, defined in Eq.~(\ref{source-term}).
The boundary conditions for the transport equation, Eq.~(\ref{transport}),
require the solution $f(T,\vec{r},t)$ to vanish at the boundary
of the diffusion zone, which is approximated by a cylinder with 
half-height $L = 1-15~\rm{kpc}$ and radius $ R = 20 ~\rm{kpc}$.
A detailed study shows that there is some degeneracy among the parameters
of the propagation model. This can be used to decrease the size of $L$
and thereby reduce the antiproton flux from gravitino decays relative to
the flux from spallation \cite{mtx02}.

The solution of the transport equation at the Solar System, 
$r=r_\odot$, $z=0$, can be formally expressed by the convolution
\begin{equation}
f(T)=\frac{1}{m_{3/2} \tau_{3/2}}
\int_0^{T{\rm max}}dT^\prime G(T,T^\prime) 
 \frac{dN(T^\prime)}{dT^\prime} ,
\label{solution}
\end{equation}
where $T_{\rm max}=m_{3/2}$ for the case of the positrons 
and $T_{\rm max}=m_{3/2}-m_p$ for the antiprotons. The 
Green's function $G(T,T^\prime)$ encodes all the information
about astrophysics (such as the details of the halo profile and 
the propagation of antiparticles in the Galaxy), while the
remaining part depends on gravitino properties. Analytical
and numerical expressions for the Green's function for the case 
of positrons and antiprotons can be found in~\cite{it08}.

Finally, the flux of primary antiparticles at the Solar System
from gravitino decay is given by:
\begin{equation}
\Phi^{\rm{DM}}(T) = \frac{v}{4 \pi} f(T) ,
\label{flux}
\end{equation}
where $v$ is the velocity of the antimatter particle.

The calculation of the high-energy cosmic ray fluxes from
gravitino decay is hindered by a large number of uncertainties
stemming both from astrophysics, encoded in the Green's function,
and from particle physics. 
However, as we shall show below, present observations of the positron 
fraction and the antiproton flux constrain the 
parameters of the model well enough to make definite predictions on the
diffuse gamma-ray flux. These predictions will be tested by
the Fermi LAT results in the near future, thus providing a crucial test
of the scenario of decaying gravitino dark matter. 

>From the particle physics point of view, the solution to the transport
equation depends on the following unknown quantities:
the energy spectrum of positrons or antiprotons produced in the decay,
the gravitino mass and the gravitino lifetime.
The energy spectrum of positrons and antiprotons depends crucially
on the $R$-parity breaking interactions of the gravitino. It was shown
in Section 2 that when the gravitino mass is large,
the branching ratios for the dominant decay channels are predicted to be 
$\sum_i \br(\psi_{3/2}\rightarrow h\nu_i)\simeq 1/4$, 
$\sum_i\br(\psi_{3/2}\rightarrow Z\nu_i)\simeq 1/4$, 
$\sum_i\br(\psi_{3/2}\rightarrow W^{\pm}e^{\mp}_i)\simeq 1/2$,
while the branching ratio for $\psi_{3/2}\rightarrow \gamma \nu$
is predicted to be much smaller. Therefore, in this limit the injection
spectrum of antiprotons from gravitino decay is fairly model-independent, 
being just a function of the gravitino mass. 
This is not the case, however, for the energy spectrum of positrons,
since the flavour composition of the final state
depends on the flavour structure of the $R$-parity breaking couplings,
which cannot be predicted without invoking a model of flavour.

The observation by PAMELA of an excess in the positron fraction
at energies extending at least until 100 GeV implies a lower
bound on the gravitino mass of $\sim 200$ GeV if this positron excess 
is interpreted in terms of gravitino decays. Besides, 
as shown in \cite{bes08}, there exists a theoretical upper
bound on the gravitino mass of $\sim 600$ GeV in supergravity models
with universal gaugino masses at the GUT scale, stemming from
the requirement of successful thermal leptogenesis, yielding a 
relatively narrow range for the gravitino mass, $m_{3/2}\simeq 200-600$ GeV. 
If the positron excess observed by PAMELA is unrelated to gravitino decays, 
the gravitino mass may be as low as $\sim 5$ GeV without yielding overclosure 
of the Universe \cite{bcx07}.

\subsection{Constraints from the antiproton flux}

The interstellar antiproton flux from gravitino decay can 
be calculated from Eq.~(\ref{solution}) using the corresponding
Green's function. However, this is not the antiproton flux
measured by antiproton experiments, which is affected at low energies
by solar modulation. Under the force field 
approximation~\cite{solar-modulation}, the antiproton flux
at the top of the Earth's atmosphere is related 
to the interstellar antiproton flux~\cite{perko}
by the simple relation
\begin{equation}
\Phi_{\bar p}^{\rm TOA}(T_{\rm TOA})=
\left(
\frac{2 m_p T_{\rm TOA}+T_{\rm TOA}^2}{2 m_p T_{\rm IS}+T_{\rm IS}^2}
\right)
\Phi_{\bar p}^{\rm IS}(T_{\rm IS})\ ,
\end{equation}
where $T_{\rm IS}=T_{\rm TOA}+\phi_F$, with
$T_{\rm IS}$ and $T_{\rm TOA}$ being the antiproton kinetic energies 
at the heliospheric boundary and at the top of the Earth's atmosphere,
respectively, and $\phi_F$ being the solar modulation parameter,
which varies between 500 MV and 1.3 GV over the eleven-year solar
cycle. Since experiments are usually undertaken near
solar minimum activity, we shall choose $\phi_F=500$ MV 
for our numerical analysis in order to compare our predicted flux with 
the collected data.

\begin{figure}
\begin{center}
\includegraphics[scale=0.85]{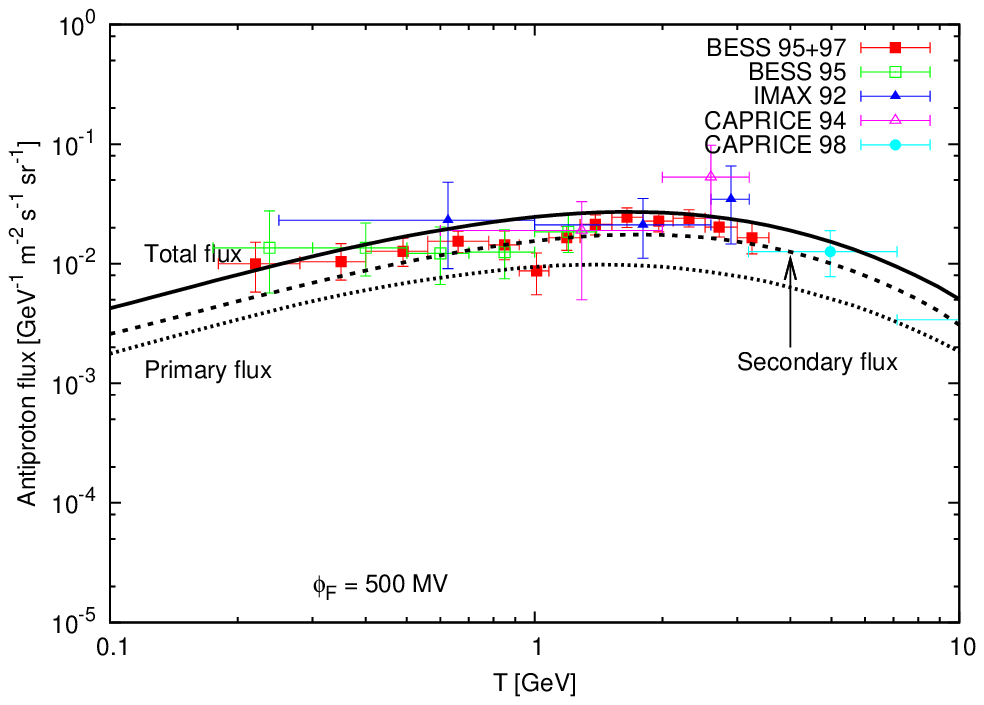}\\
\includegraphics[scale=0.85]{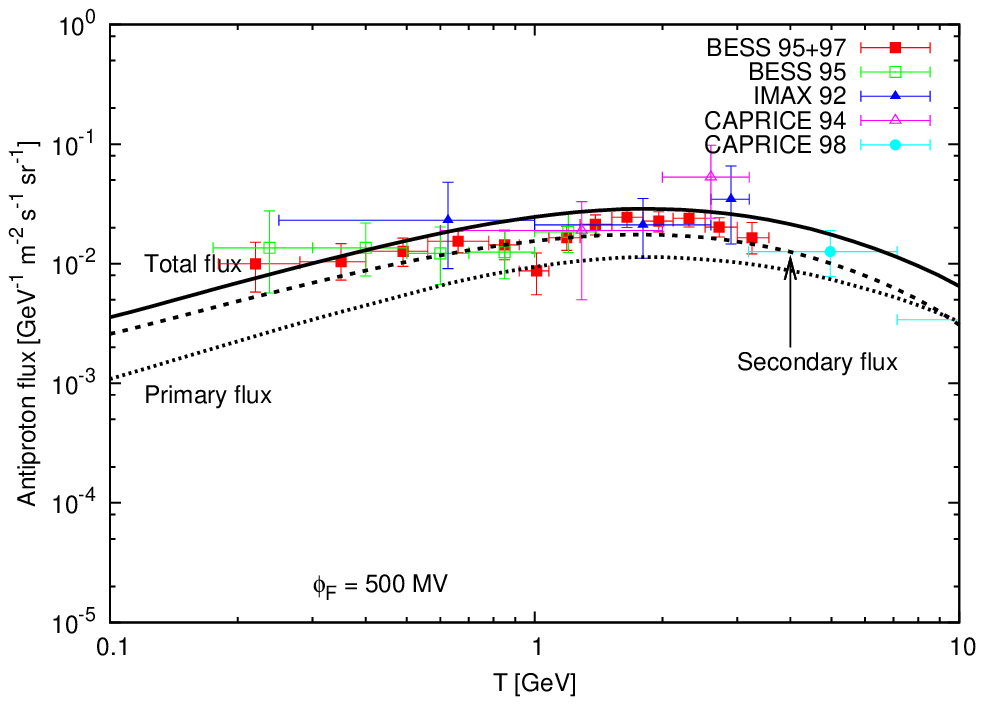}\\
\includegraphics[scale=0.85]{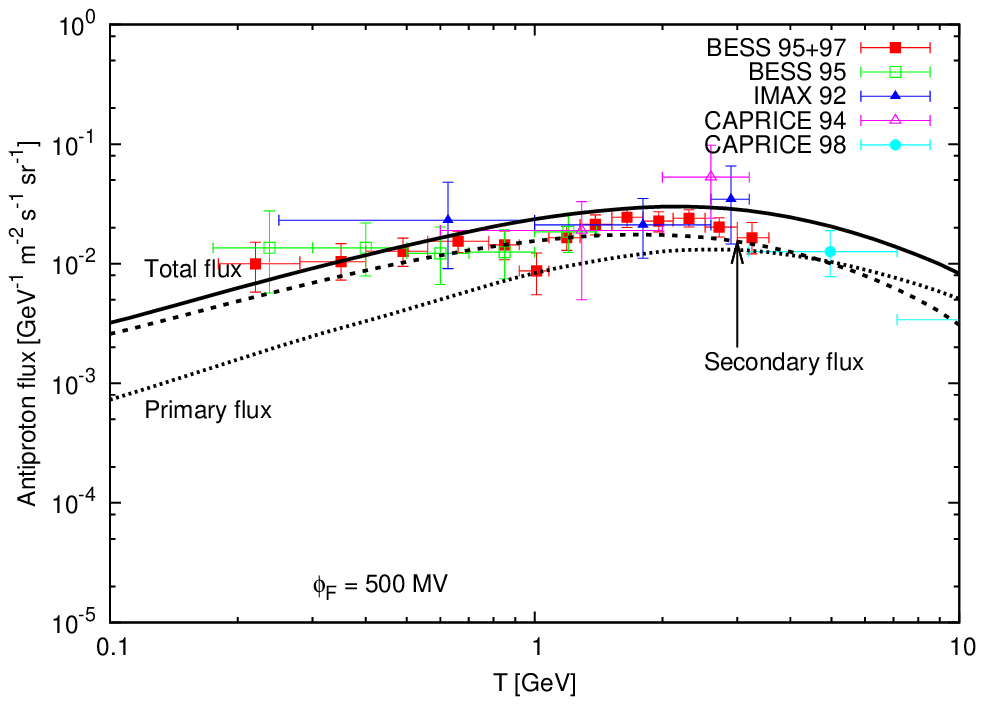}
\end{center}
\caption{
Antiproton fluxes for $m_{3/2} = 200,~400,~600~\mathrm{GeV}$ in the 
MED set of propagation parameters that saturate the antiproton 
overproduction bound (see text). Dotted lines: antiproton 
flux from gravitino decays, dashed lines: secondary antiproton flux from 
spallation in the case of minimal nuclear cross sections, solid lines: 
total antiproton flux. The gravitino lifetimes are 
$\tau_{3/2} = 7 \times 10^{26}~\text{s},~3 \times 10^{26}~\text{s}$ and 
$1.5 \times 10^{26}~\text{s}$, respectively.
}
\label{fig:pbar}
\end{figure}
 
As discussed in \cite{it08}, the calculation
of the antiproton flux from gravitino decay suffers from 
uncertaintes in the determination of the physical parameters in 
the propagation of charged cosmic rays in the diffusive halo, 
leading to uncertainties in the magnitude of fluxes as large
as two orders of magnitude at the energies relevant
for present antiproton experiments. The requirement that the total 
antiproton flux from gravitino decay be consistent with 
measurements gives a lower bound on the gravitino mass
which strongly depends on the choice of the halo model.
In the following we shall adopt the MED propagation model,
which provides the best fit to the B/C ratio and measurements of 
flux ratios of radioactive cosmic-ray species~\cite{Maurin:2001sj}.

In order to determine the maximally allowed exotic contribution to
the total antiproton flux, a precise knowledge of
the secondary flux of antiprotons from spallation of high-energy 
cosmic rays on the protons and helium nuclei in the interstellar medium 
is necessary. 
Unfortunately, the determination
of this secondary flux is also subject to uncertainties. First,
the choice of the propagation model can change the prediction
of the secondary flux by $10-20$\%. More importantly, the
uncertainty in the nuclear cross sections for p-p, p-He,
He-p and He-He collisions can change the prediction of the
secondary flux by $22-25$\% above or below the central 
value~\cite{Donato:2001ms}.

A conservative upper bound on the antiproton flux from gravitinos is obtained
by demanding that the total flux is not larger than the theoretical uncertainty
band of the MED propagation model. This means that a `minimal' dark matter
lifetime for the MED model can be defined by a scenario where the 
secondary antiproton flux from spallation is 25\% smaller
than the central value, due to a putative overestimation of
the nuclear cross sections, and the total antiproton flux saturates
the upper limit of the uncertainty band which stems from astrophysical 
uncertainties discussed above. This amounts to
the requirement that the antiproton flux from gravitino decays 
should not exceed $\sim 50\%$ of the central value of secondary flux 
from spallation. 

Using the above prescription we find the following lower
bounds for the gravitino lifetime: 
\begin{equation}\label{minlife}
\tau^\mathrm{min}_{3/2}(200) \simeq 7\times 10^{26}~\mathrm{s}  , \quad
\tau^\mathrm{min}_{3/2}(400) \simeq 3\times 10^{26}~\mathrm{s}  , \quad
\tau^\mathrm{min}_{3/2}(600) \simeq 1.5\times 10^{26}~\mathrm{s}  ,
\end{equation}
where the numbers in parentheses correspond to the gravitino masses
$m_{3/2} = 200,~400~\mathrm{and}~600~\mathrm{GeV}$, respectively. 
The corresponding antiproton fluxes from gravitino decay, the
secondary antiproton flux from spallation and the total antiproton fluxes
are shown in Figure~\ref{fig:pbar} together with the experimental measurements
by BESS, IMAX and WiZard/CAPRICE, and the uncertainty band 
from the nuclear cross sections in the MED propagation model.
The minimal lifetimes (\ref{minlife}) can be compared with the gravitino
lifetimes needed to explain the PAMELA positron fraction excess, which
will be discussed in the next section.

\subsection{Comparison with electron/positron fluxes}

Using the procedure described in the previous section it is
straightforward to calculate the positron flux at Earth from
gravitino decay. we shall adopt for definiteness
the MED propagation model~\cite{Maurin:2001sj}, characterised 
by $\delta=0.70$, $K_0=0.0112\,{\rm kpc}^2/{\rm Myr}$,
$L=4\,{\rm kpc}$, $V_c=12\,{\rm km}/{\rm s}$.  Note that
the sensitivity of the positron fraction 
to the propagation model is fairly mild at the energies where
the excess is observed, since these positrons were produced
within a few kiloparsecs from the Earth and barely suffered
the effects of diffusion.

\begin{figure}
\begin{center}
\includegraphics[scale=0.9]{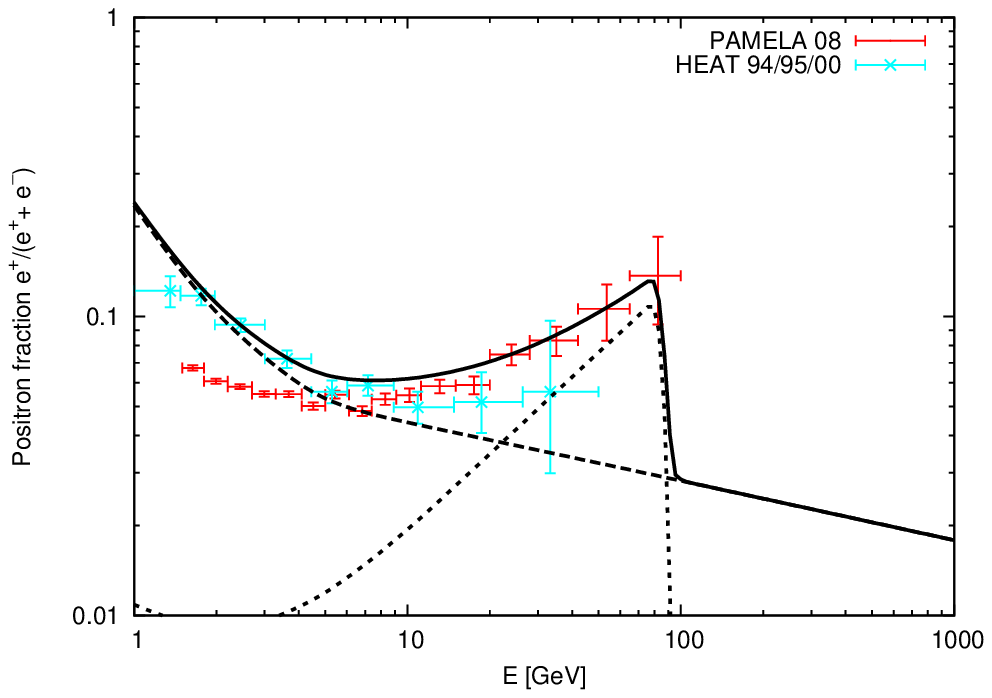}\\
\includegraphics[scale=0.9]{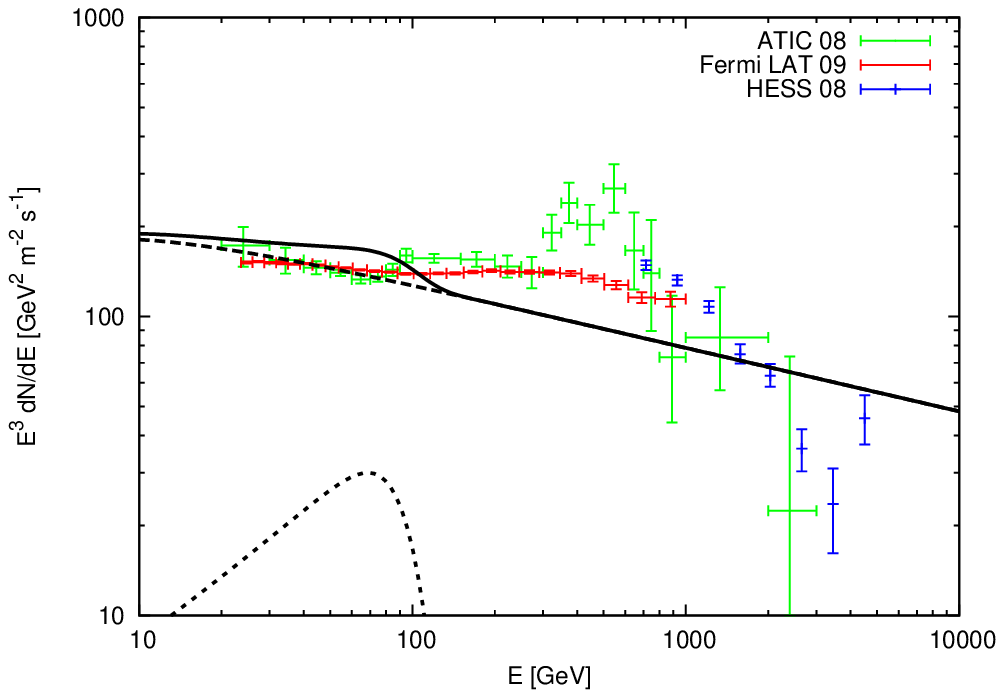}
\end{center}
\caption{
Contribution from dark matter decay to the positron fraction and the total
electron $+$ positron flux, compared with data from PAMELA and HEAT, and
ATIC, Fermi LAT and HESS, respectively; 
$m_{3/2} = 200~\text{GeV},~\tau_{3/2} = 3.2 \times 10^{26}~\text{s}$, for
$W^{\pm}l^{\mp}$ decays pure electron flavour is assumed. 
The ``Model 0'' background is used, and for comparison with 
Fermi-LAT data 25\% energy resolution is taken into account.
}
\label{fig:pamela}
\end{figure}

To compare the predictions
to the PAMELA results, we shall calculate the positron fraction,
defined as the ratio of the positron flux to the combined positron 
and electron flux, $\Phi_{e^+}/(\Phi_{e^+}+\Phi_{e^-})$. 
For the background fluxes of primary and secondary electrons, 
as well as secondary positrons, we extract the fluxes from 
``Model 0'' presented by the Fermi
collaboration in \cite{gxx09}, which fits well the low energy data points of
the total electron plus positron flux and the positron fraction,
and is similar to the MED model for energies
above a few GeV \cite{Delahaye:2008ua}.
Then, the positron fraction reads
\begin{equation}
{\rm PF}(T) = \frac{\Phi_{e^+}^{\rm{DM}}(T) + \Phi_{e^+}^{\rm{bkg}}(T)}
{\Phi_{e^+}^{\rm{DM}}(T) + \Phi_{e^+}^{\rm{bkg}}(T) +\Phi_{e^-}^{\rm{DM}}(T)
+ k \;  \Phi_{e^-}^{\rm{bkg}}(T) },
\label{PF}
\end{equation}
where $k = {\cal O}(1)$ is the normalization of the 
astrophysical contribution to the primary electron flux,
which is chosen to provide a qualitatively good fit to the data.

\begin{figure}
\begin{center}
\includegraphics[scale=0.7]{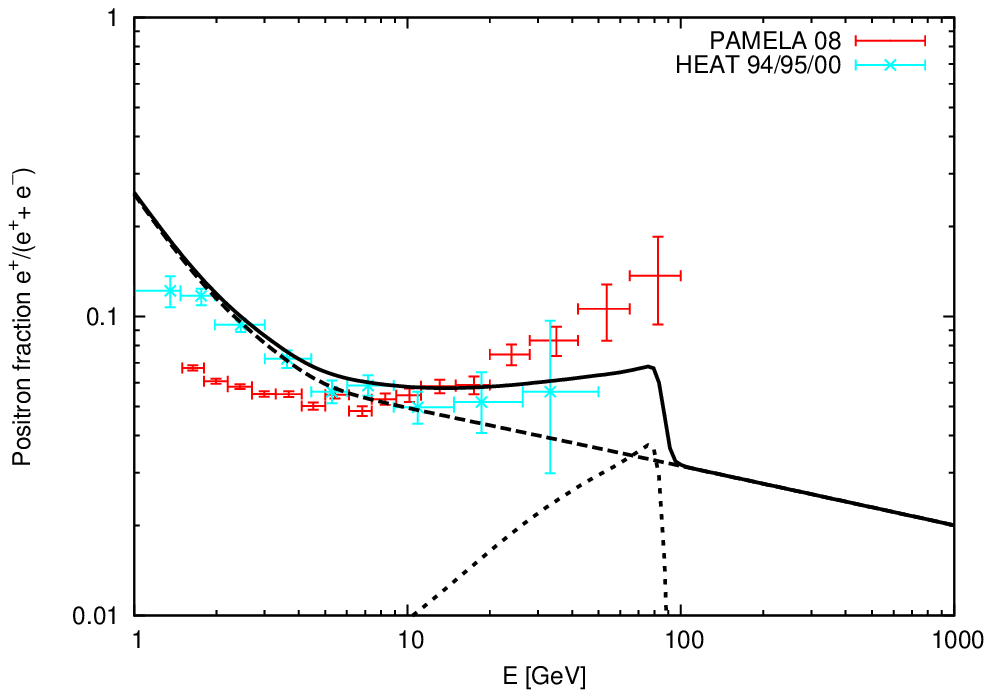}
\includegraphics[scale=0.7]{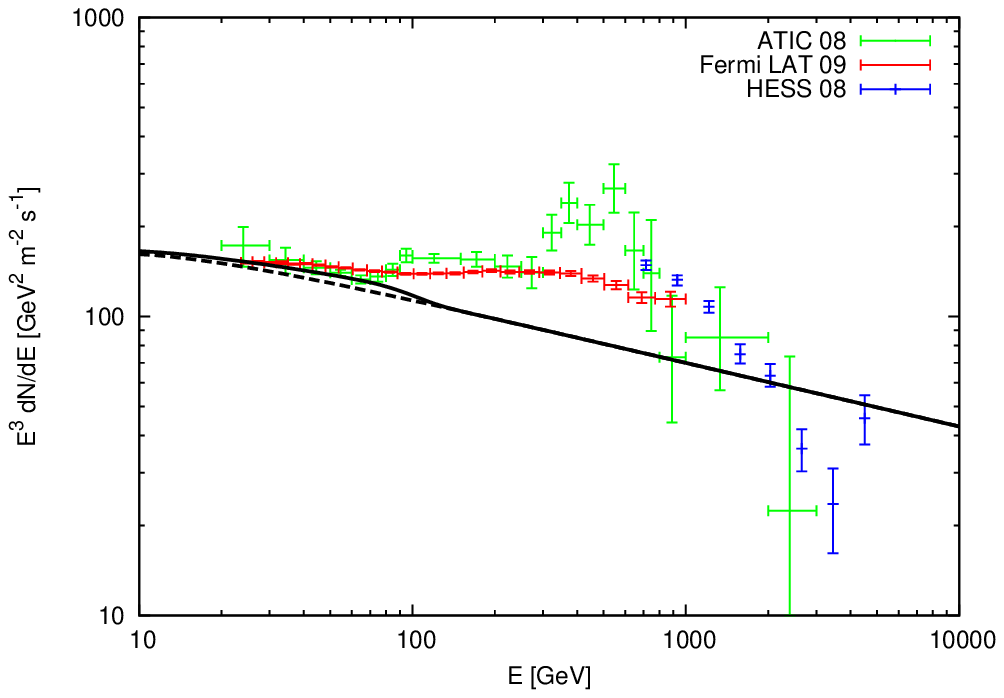}\\
\includegraphics[scale=0.7]{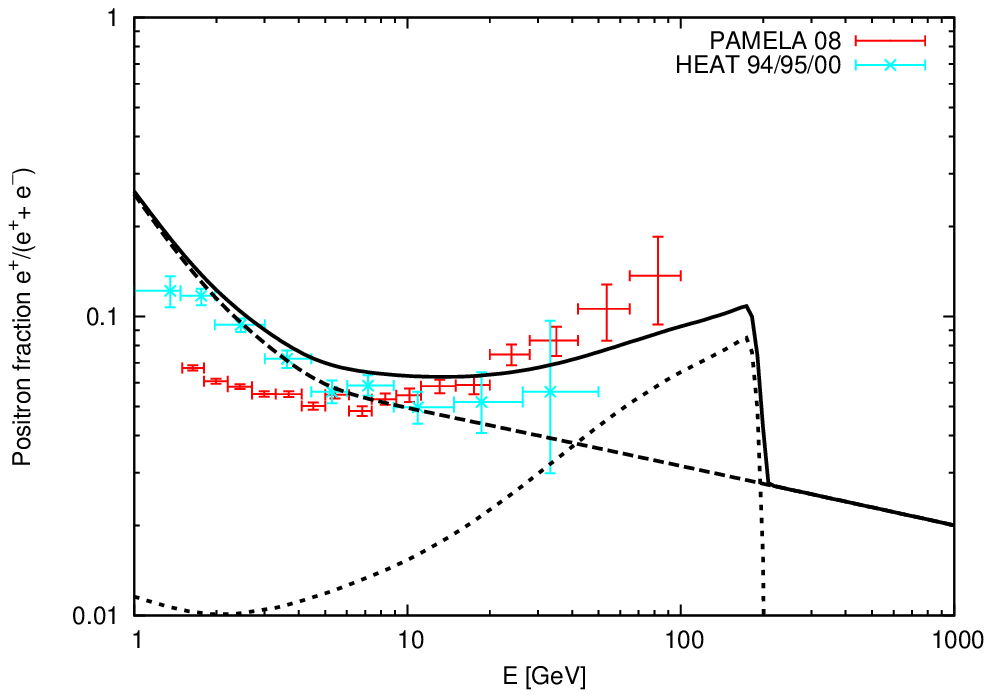}
\includegraphics[scale=0.7]{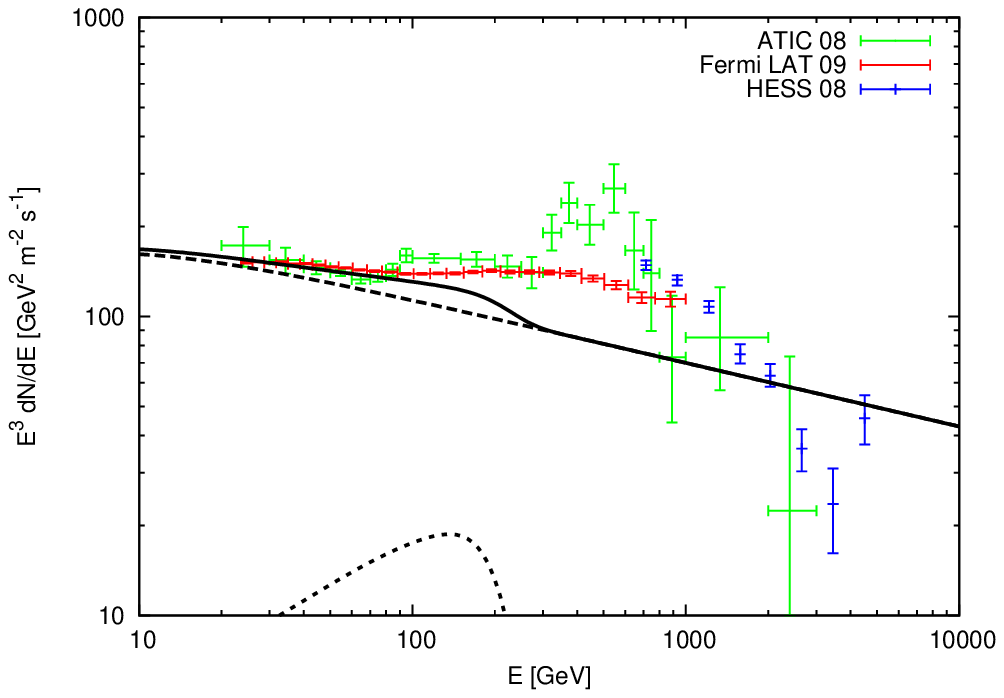}\\
\includegraphics[scale=0.7]{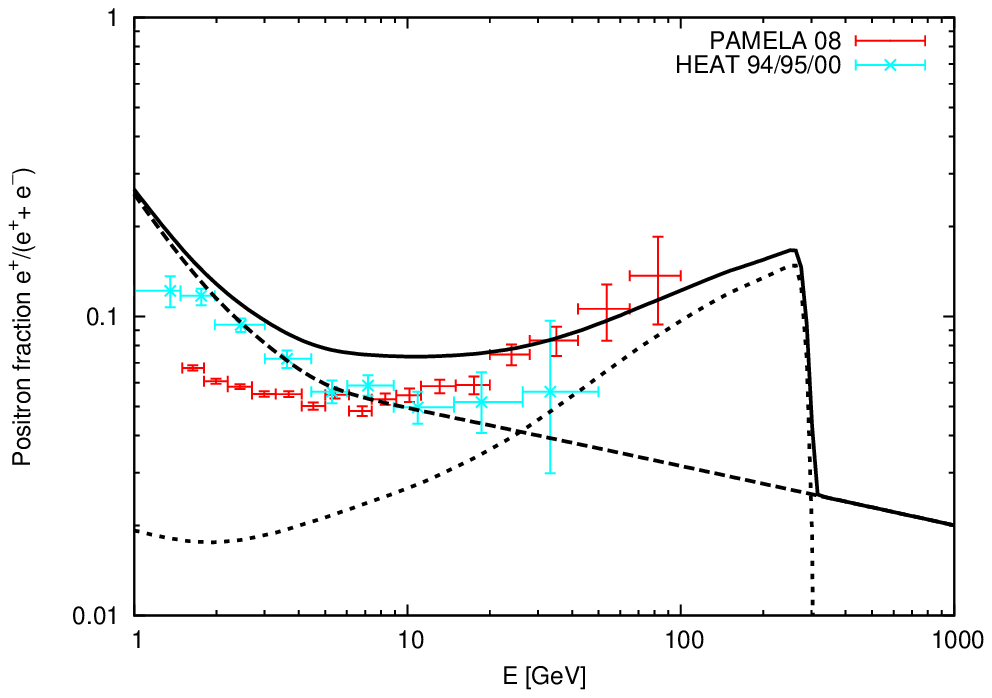}
\includegraphics[scale=0.7]{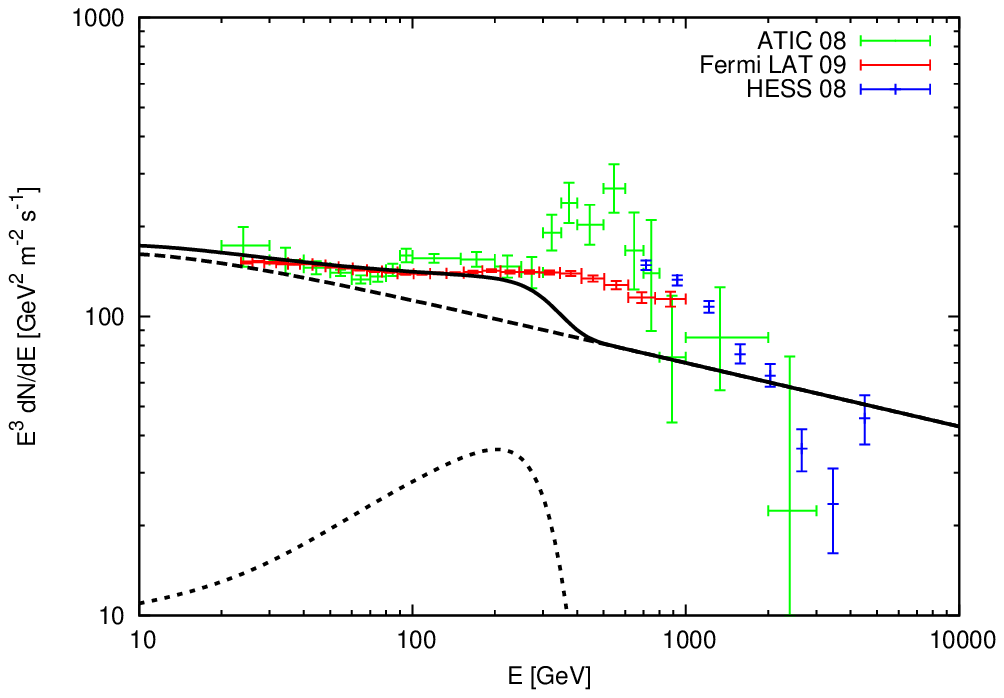}
\end{center}
\caption{
Contribution from dark matter decay to the positron fraction and the total
electron $+$ positron flux, compared with data from PAMELA and HEAT, and
ATIC, Fermi LAT and HESS, respectively; 
$m_{3/2} = 200,~400,~600~\text{GeV}$ with the minimal lifetimes (\ref{minlife}) (top to bottom); for $W^{\pm}l^{\mp}$ decays democratic flavour dependence is 
assumed. The ``Model 0'' background is used, and for comparison with 
Fermi LAT data 25\% energy resolution is taken into account.
}
\label{totaldem}
\end{figure}

We now discuss the hypothesis that the PAMELA positron excess 
is due to gravitino dark matter decay. For this to be the case, the 
gravitino mass must be at least 200 GeV. As discussed above, the branching 
ratios into Standard Model particles will be essentially fixed for gravitino 
masses of a few hundred GeV. The decay $\psi_{3/2} \rightarrow W^\pm \ell^\mp$ 
then has a branching ratio of $\sim$ 50\%, and the hard leptons that are 
directly produced in these decays may account for the rise in the 
positron fraction if a significant fraction of these leptons has 
electron or muon flavour. 

Consider first the extreme case that the decays occur purely into electron 
flavour. For $m_{3/2}=200~\mathrm{GeV}$, the PAMELA excess can then be 
explained for the gravitino lifetime 
$\tau^e_{3/2}(200) \simeq 3.2\times 10^{26}~\mathrm{s}$,
as illustrated by Figure~\ref{fig:pamela}. 
Note that this lifetime is a factor 2 smaller than the minimum lifetime given
in (\ref{minlife}), which we obtained from the antiproton constraint. 
In other words, an interpretation 
of the PAMELA excess in terms of gravitino decays is incompatible with 
the MED set of propagation parameters once antiprotons are taken into 
account. Nevertheless, the MIN model and other sets of parameters that 
yield intermediate values for the antiproton flux can easily be 
compatible with both the positron fraction and the antiproton-to-proton 
ratio observed by PAMELA. The situation is very similar for 
$m_{3/2}=400~\mathrm{and}~600~\mathrm{GeV}$.

Figure~\ref{fig:pamela} also shows the predicted total electron $+$ positron 
flux together with the results from Fermi and ATIC. Obviously, the ``Model 0''
presented by the Fermi collaboration in \cite{gxx09}, cannot account for the present data, and the contribution
from gravitino decays makes the discrepancy even worse. In particular, the
data show no spectral feature expected for decaying dark matter. On the
other hand, gravitino decays may very well be consistent with the measured
total electron $+$ positron flux once the background is appropriately
adjusted. This is evident from Figure~\ref{totaldem} where the contribution
from gravitino decays is shown in the theoretically well motivated case
of flavour democratic decays. The figure also illustrates that, depending
on the gravitino mass, the dark matter contribution to the PAMELA excess
can still be significant.

\begin{figure}
\begin{center}
\includegraphics[scale=0.7]{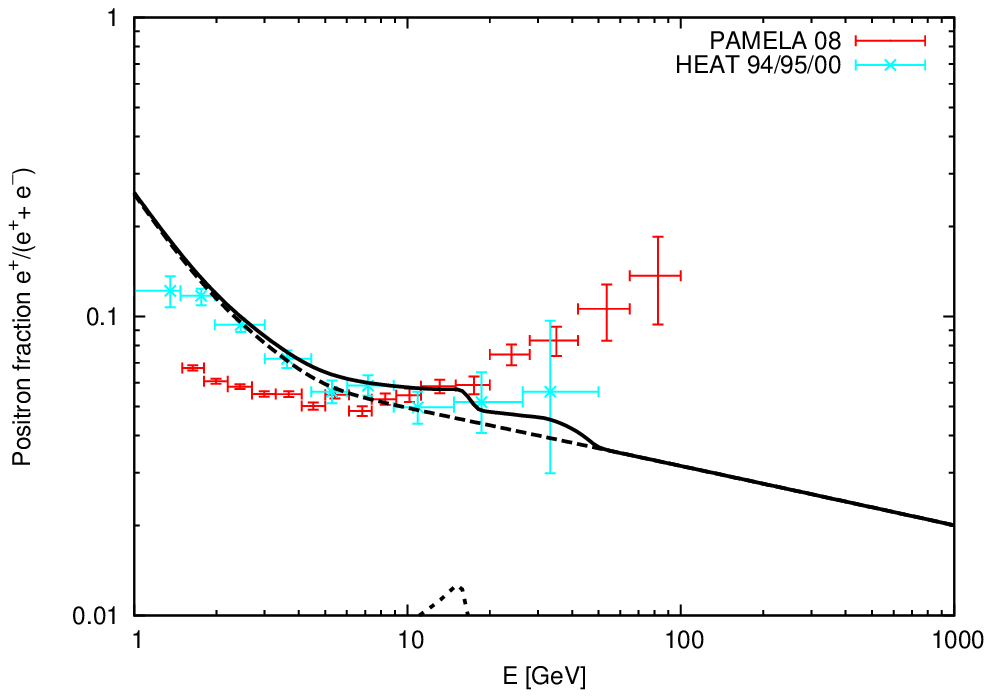}
\includegraphics[scale=0.7]{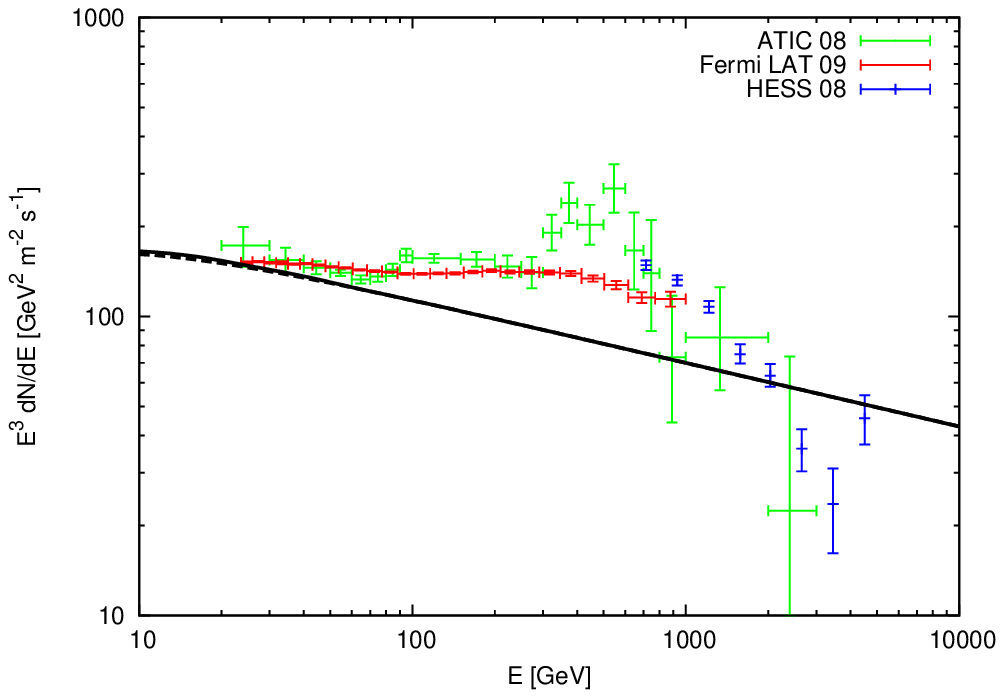}
\end{center}
\caption{Contribution of gravitino decays to positron fraction and
total electron + positron flux for $m_{3/2} = 100$~GeV and
$\tau_{3/2} = 1\times 10^{27}$~s.}
\label{fig:pam100}
\end{figure}

An obvious possibility is that both, the total electron + positron flux and
the positron fraction, are dominated by astrophysical sources. For instance,
for the gravitino mass $m_{3/2} = 100$~GeV we obtain from the antiproton flux
constraint $\tau_{3/2}^{\rm min}(100) \simeq 1\times 10^{27}$~s. As 
Figure~\ref{fig:pam100} demonstrates, the contribution from gravitino decays
to the total electron + positron flux and positron fraction is indeed 
negligible. Nevertheless, as we shall see in the following section, the 
dark matter contribution to the gamma-ray flux can be sizable.

\section{Predictions for the diffuse gamma-ray flux}

The total gamma-ray flux from gravitino dark matter
decay receives two main contributions. The first
one stems from the decay of gravitinos in the Milky Way halo,
\begin{equation}
\left[E^2 \frac{dJ}{dE}\right]_{\text{halo}} = 
\frac{2 E^2}{m_{3/2}} \frac{dN_\gamma}{dE} 
\frac{1}{8 \pi \tau_{3/2}} \int_\text{los} \rho_\text{halo}(\vec{l}) d\vec{l}
\;,
\label{halo-flux}
\end{equation}
where $dN_\gamma/dE$ is the
gamma-ray spectrum produced in the gravitino decay.
The integration extends over the line of sight, so the halo contribution
has an angular dependence on the direction of observation.
 
In addition to the cosmological contribution, the total gamma-ray flux also 
receives a contribution from the decay of gravitinos at cosmological
distances, giving rise to a perfectly isotropic extragalactic 
diffuse gamma-ray background. The flux 
received at the Earth with extragalactic origin is given by
\begin{equation}
\left[E^2 \frac{dJ}{dE}\right]_{\text{eg}} =
 \frac{2 E^2}{m_{3/2}} C_\gamma \int_1^\infty dy 
\frac{dN_\gamma}{d(E y)} \frac{y^{-3/2}}
{\sqrt{1 + \Omega_\Lambda/\Omega_M y^{-3}}} ,
\label{extgal-flux}
\end{equation}
where $y=1+z$, $z$ being the redshift, and
\begin{equation}
C_\gamma = \frac{\Omega_{3/2} \rho_c}{8 \pi \tau_{3/2} H_0 \Omega_M^{1/2}}
\simeq 10^{-6}\ (\text{cm}^2\text{s}~\text{str})^{-1} \text{GeV}
\left(\frac{\tau_{3/2}}{ 10^{27}\ \mbox{s}} \right)^{-1} .
\end{equation}
Here, $ \Omega_{3/2} $, $ \Omega_M $ and $ \Omega_{\Lambda} $ are the 
gravitino, matter and cosmological constant density parameters, 
respectively, $ \rho_c $ is the critical 
density and $ H_0 $ the present value of the Hubble parameter.

\begin{figure}
\begin{center}
\includegraphics[scale=0.85]{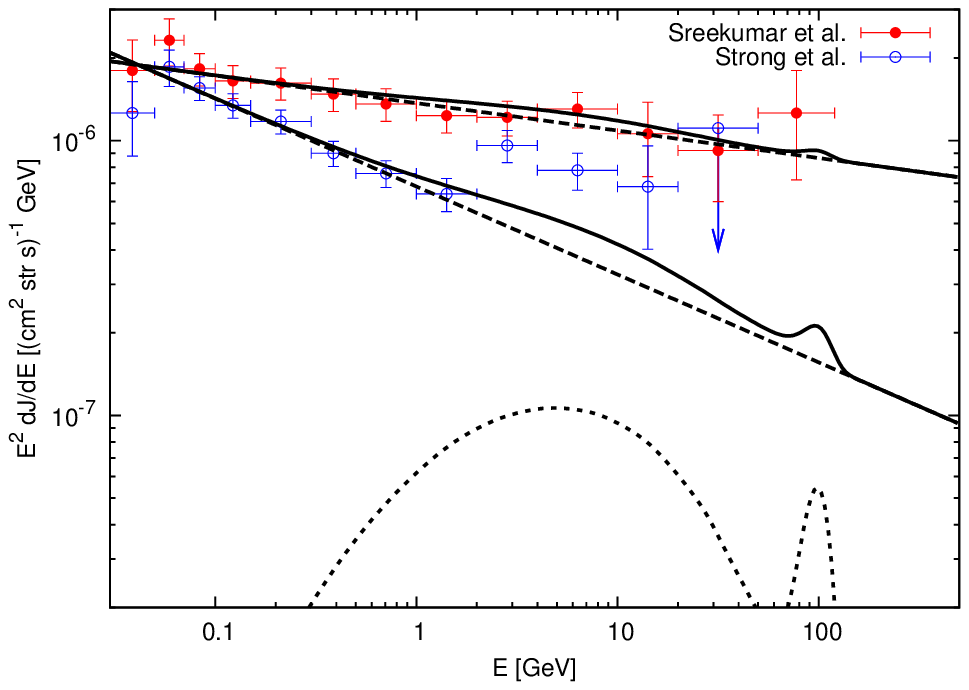}\\
\includegraphics[scale=0.85]{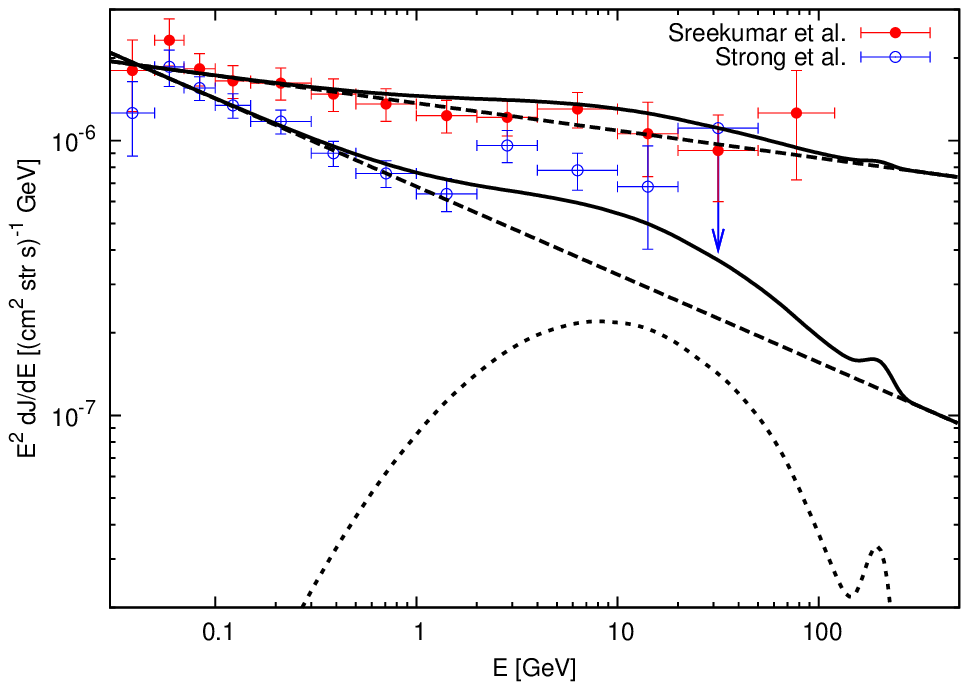}\\
\includegraphics[scale=0.85]{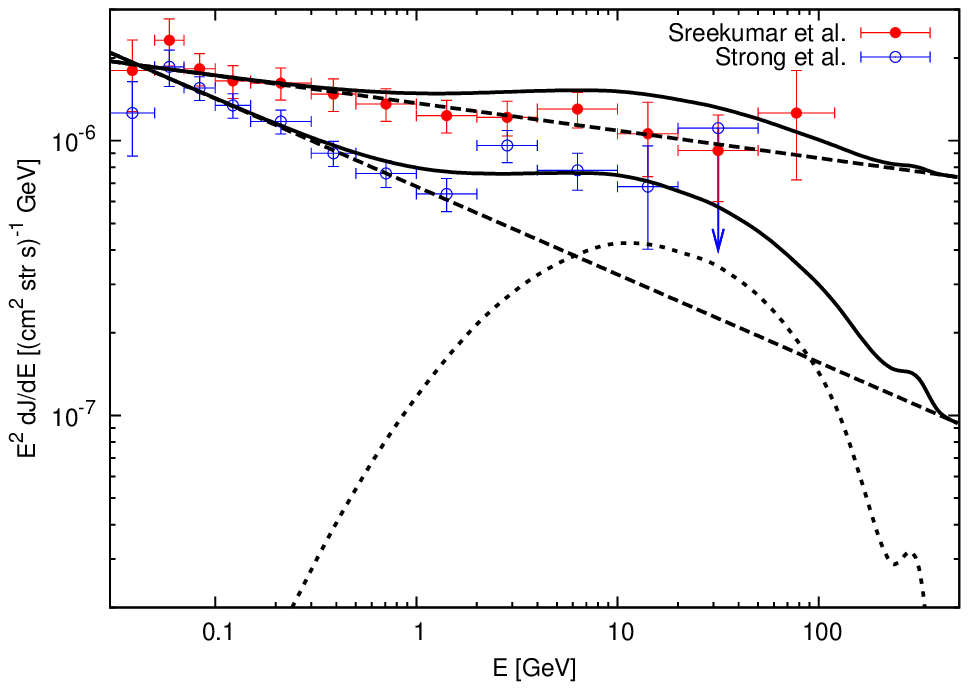}
\end{center}
\caption{
Predicted gamma-ray flux for $m_{3/2} = 200,~400,~600~\mathrm{GeV}$ for the 
minimal lifetimes (\ref{minlife}). We assume decays 
purely into electron flavour here. We show both the background 
obtained by Sreekumar et al. as well as the background obtained by 
Strong, Moskalenko and Reimer.
}
\label{fig:gamma-min}
\end{figure}

The cosmological contribution is numerically smaller than the halo 
contribution. Moreover, the flux of gamma-rays of cosmological
origin is attenuated by the electron-positron pair production 
on the extragalactic background light emitted by galaxies in the
ultraviolet, optical and infrared frequencies~\cite{Gould:1967zzb}.
However, the flux of gamma-rays originating from the decay of dark matter
particles in the halo is barely attenuated by pair production on the Galactic 
interstellar radiation field at energies below
10 TeV~\cite{Moskalenko:2005ng}. Thus, the total flux is dominated
by the halo component, yielding a slightly anisotropic gamma-ray 
flux~\cite{bbx07} which is compatible with the EGRET observations~\cite{it07}.
A detailed study of the prospects of detecting this anisotropy with the 
Fermi LAT is beyond the scope of this paper and will be presented 
in more generality elsewhere~\cite{ITW}.

The gravitino decay produces a continuous spectrum of gamma-rays
which is determined by the fragmentation of the Higgs boson
and the weak gauge bosons. On the other hand, there exists a 
gamma-ray line at the endpoint of the spectrum with an intensity
which is model-dependent.\footnote{We neglect in our analysis the 
contribution
to the gamma-ray flux from inverse Compton scattering of high energy
electrons and positrons on the interstellar radiation field. We estimate
that this contribution is peaked at energies smaller than 0.1 GeV and
has an intensity $E^2 \frac{dJ}{dE}\lsim {\cal O}(10^{-7})
(\text{cm}^2~\text{str}~\text{s})^{-1}~\text{GeV}$~\cite{Ishiwata:2009dk}, 
thus giving a negligible contribution to the total flux, which is 
constrained by EGRET to be $E^2 \frac{dJ}{dE}\sim 10^{-6}
(\text{cm}^2~\text{str}~\text{s})^{-1}~\text{GeV}$ at $E=0.1$GeV.
This contribution, however, can be sizable for larger dark matter
masses~\cite{Ishiwata:2009dk,Ibarra:2009dr}.}
For our numerical analysis we shall
use the typical branching ratio in this channel derived in Section~2,
\begin{equation}
\br(\psi_{3/2}\rightarrow \nu\gamma) = 
0.02 \left(\frac{200~\mathrm{GeV}}{m_{3/2}}\right)^2 , \nonumber
\end{equation}
for gravitino masses in the range from $100 - 600$~GeV.
In Figure~\ref{fig:gamma-min} the predicted diffuse gamma-ray
flux is shown for $m_{3/2} = 200,~400,~600~\mathrm{GeV}$ and the
respective lower bounds (\ref{minlife}) on the gravitino lifetime. These 
spectra correspond to upper bounds on the 
signal in gamma-rays that can be expected from gravitino dark matter decay. 

For comparison, we show two sets of data points obtained from the 
EGRET measurements of the diffuse extragalactic gamma-ray background 
using different models of the Galactic foreground emission. 
An analysis by Strong, Moskalenko and Reimer using a model, optimised to 
better simulate the Galactic diffuse emission, revealed a power law 
behaviour below 1 GeV, with an intriguing deviation from a power law 
above 1 GeV \cite{smr04}.
For our present analysis, we shall show both sets of results as 
the status of the extragalactic background is currently unclear. 
For the background obtained by Moskalenko, Strong and Reimer, 
the extragalactic component is described by the power law \cite{smr04}
\begin{equation}
\left[E^2 \frac{dJ}{dE}\right]_\text{bg}=6.8\times 10^{-7}
\left(\frac{E}{\rm GeV}\right)^{-0.32}
(\text{cm}^2~\text{str}~\text{s})^{-1}~\text{GeV} . 
\end{equation}
The earlier analysis by Sreekumar et al led to
a less steep background \cite{sxx97},
\begin{equation}
\left[E^2 \frac{dJ}{dE}\right]_\text{bg}=1.37\times 10^{-6}
\left(\frac{E}{\rm GeV}\right)^{-0.1}
(\text{cm}^2~\text{str}~\text{s})^{-1}~\text{GeV} . 
\end{equation}
For comparison with the data points, we have averaged the 
slightly anistropic halo signal over the whole sky,
excluding a band of $\pm 10^\circ$ around the Galactic 
disk.\footnote{ The halo signal would have a larger degree of anisotropy
if the dark matter halo is not completely uniform but presents
substructures, as suggested by N-body simulations of Milky-Way-size
galaxies. 
In the present calculation
we are interested in the average flux in the whole sky excluding
the Galactic disk, which depends on the total
amount of dark matter in this region and not on the way it is
distributed. Therefore, for our purposes it is a good approximation
to neglect substructures and to assume a smooth dark matter halo profile.}
We have conservatively used an energy resolution $\sigma(E)/E=15\%$ 
as quoted by Fermi \cite{Fermi-performance}.

It is remarkable that for both choices of the extragalactic background, 
the antiproton constraint allows for a sizable deviation from 
a power law background if the gravitino mass is above $200$ GeV. 
Therefore, if such a deviation with the proper angular dependence 
is observed by Fermi LAT, the scenario of gravitino dark
matter will gain support.
Furthermore, the existence of a gamma-ray line at the end of the spectrum 
is predicted, with an intensity
that, as discussed in Section 2, depends on the model of $R$-parity
breaking. This line could be observed
by Fermi LAT in the diffuse gamma-ray background, but also
by the ground-based Cherenkov telescopes MAGIC, HESS
or VERITAS in galaxies such as M31~\cite{bbx07}.
For smaller gravitino masses the gamma-ray line becomes more prominent
whereas the contribution to the continuous part of the spectrum decreases.
This is illustrated in Figure~6 for $m_{3/2} = 100$~GeV.

\begin{figure}
\begin{center}
\includegraphics[scale=0.9]{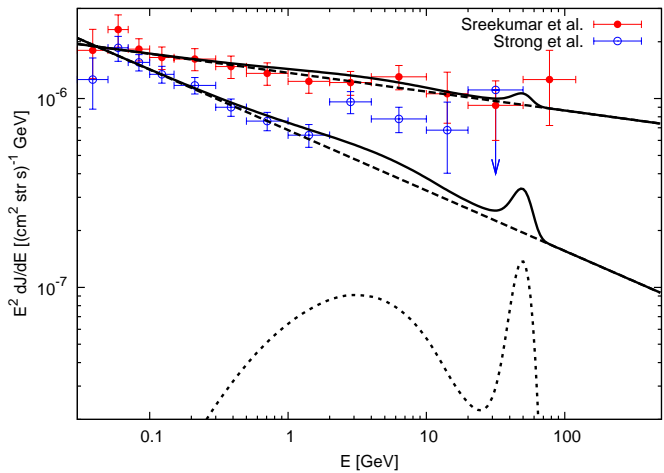}
\end{center}
\caption{
Predicted gamma-ray flux for $m_{3/2} = 100~\mathrm{GeV}$ and
$\tau_{3/2} = 1\times 10^{27}$~s. 
We show both the background obtained by 
Sreekumar et al. as well as the background obtained by Strong, Moskalenko 
and Reimer.
}
\label{fig:gamma}
\end{figure}

The observation of the discussed features in the diffuse gamma-ray 
spectrum might, if interpreted as the result of gravitino
decay, open the exciting 
possibility of constraining the reheating temperature of the Universe.
More concretely, the thermal relic abundance of gravitinos is given by 
\cite{bbb00,ps06,bes08}
\begin{equation}
\Omega^{\text{th}}_{3/2} h^2\simeq 0.5
    \left(\frac{T_R}{10^{10}\,\text{GeV}}\right)
    \left(\frac{100 \,\text{GeV}}{m_{3/2}}\right)
    \left(\frac{m_{\widetilde g}}{1\,\text{TeV}}\right)^2\;.
\label{relic-abundance}
\end{equation}
Therefore, imposing that the thermal abundance of gravitinos should
not be larger that the total dark matter abundance, 
the measurement of the gravitino mass by Fermi LAT
and the measurement of the gluino mass at the LHC imply
the following upper bound
on the reheating temperature of the Universe:
\begin{equation}
  T_R \lsim 2\times 10^{9}\,\text{GeV}
    \left(\frac{\Omega_{3/2} h^2}{0.1}\right)
    \left(\frac{100 \,\text{GeV}}{m_{3/2}}\right)^{-1}
    \left(\frac{m_{\widetilde g}}{1\,\text{TeV}}\right)^{-2}\;,
\end{equation}
which is saturated when all the dark matter gravitinos are of 
thermal origin. This bound has impoprtant implications for
the scenario of thermal leptogenesis, which requires $T_R\gsim 10^9$ GeV,
as well as for many inflationary scenarios.

\section{Conclusions}

In supersymmetric theories with small $R$-parity breaking thermally produced
gravitinos can account for the observed dark matter, consistent with 
leptogenesis and nucleosynthesis. Gravitino decays then contribute to 
antimatter cosmic rays as well as gamma-rays. We consider gravitino masses 
below 600~GeV, which are consistent with universal boundary conditions at the
GUT scale. 

Gravitino decays into Standard Model particles can be studied in a 
model-independent way by means of an operator analysis. For sufficiently
large gravitino masses the dimension-5 operator dominates. This means
that the branching ratios into $h\nu$, $Z\nu$ and $W^{\pm}l^{\mp}$ are
fixed, except for the dependence on lepton flavour. As a consequence,
the gamma-ray flux is essentially determined once the antiproton flux is known.
On the contrary, the positron flux is model-dependent.

The gamma-ray line is controlled by the dimension-6 operator. Hence, it
is suppressed compared to the continuous gamma-ray spectrum. Its
strength is model-dependent and decreases with increasing gravitino mass. 

Electron and positron fluxes from gravitino decays, together with the 
standard GALPROP background, cannot account for both, the PAMELA 
positron fraction and the electron $+$ positron flux measured by Fermi LAT. 
For gravitino dark matter, the observed fluxes require astrophysical sources.  
However, depending on the gravitino mass and the background, the dark matter 
contribution to the electron and positron fluxes can be non-negligable.

Present data on charged cosmic rays allow for a sizable contribution of
gravitino dark matter to the gamma-ray spectrum, in particular a line
at an energy below 300~GeV. Non-observation of such a line would place
an upper bound on the gravitino lifetime, and therefore on the strength
of $R$-parity breaking, restricting possible signatures at the LHC.

\section*{Acknowledgements}

We would like to thank L. Covi, D. Horns, G. Sigl and C. Weniger for
useful discussions. The work of AI and DT was partially supported 
by the DFG Cluster of Excellence ``Origin and Structure of the Universe.''



\end{document}